# Attention-Based CNN-BiLSTM for Sleep State Classification of Spatiotemporal Wide-Field Calcium Imaging Data

Xiaohui Zhang[a†], Eric C. Landsness[b†], Hanyang Miao[b], Wei Chen[b], Michelle Tang[b], Lindsey M. Brier[c], Joseph P. Culver[c,d,e,f], Jin-Moo Lee[b,c,d], Mark A. Anastasio[a*]

[a] Department of Bioengineering, University of Illinois Urbana-Champaign, Urbana, IL 61801, USA.
[b] Department of Neurology, Washington University School of Medicine, St. Louis, MO 63110, USA.
[c] Department of Radiology, Washington University School of Medicine, St. Louis, MO 63110, USA.
[d] Department of Biomedical Engineering, Washington University School of Engineering, St. Louis, MO 63130, USA.
[e] Department of Electrical and Systems Engineering, Washington University School of Engineering, St. Louis, MO 63130, USA.
[f] Department of Physics, Washington University School of Arts and Science, St. Louis, MO 63130, USA.

† These authors contributed equally.
[*] Correspondence should be directed to: Department of Bioengineering, University of Illinois, Urbana-Champaign, IL, USA. Electronic address: maa@illinois.edu. 217-300-0314.

## Abstract
**Background:** Wide-field calcium imaging (WFCI) with genetically encoded calcium indicators allows for spatiotemporal recordings of neuronal activity in mice. When applied to the study of sleep, WFCI data are manually scored into the sleep states of wakefulness, non-REM (NREM) and REM by use of adjunct EEG and EMG recordings. However, this process is time-consuming, invasive and often suffers from low inter- and intra-rater reliability. Therefore, an automated sleep state classification method that operates on spatiotemporal WFCI data is desired.
**New Method:** A hybrid network architecture consisting of a convolutional neural network (CNN) to extract spatial features of image frames and a bidirectional long short-term memory network (BiLSTM) with attention mechanism to identify temporal dependencies among different time points was proposed to classify WFCI data into states of wakefulness, NREM and REM sleep.
**Results:** Sleep states were classified with an accuracy of 84% and Cohen's $\kappa$ of 0.64. Gradient-weighted class activation maps revealed that the frontal region of the cortex carries more importance when classifying WFCI data into NREM sleep while posterior area contributes most to the identification of wakefulness. The attention scores indicated that the proposed network focuses on short- and long-range temporal dependency in a state-specific manner.
**Comparison with Existing Method:** On a 3-hour WFCI recording, the CNN-BiLSTM achieved a $\kappa$ of 0.67, comparable to a $\kappa$ of 0.65 corresponding to the human EEG/EMG-based scoring.
**Conclusions:** The CNN-BiLSTM effectively classifies sleep states from spatiotemporal WFCI data and will enable broader application of WFCI in sleep.

## 1. Introduction

Wide-field calcium imaging (WFCI) with genetically encoded calcium indicators is a powerful imaging technique that allows for simultaneous recordings of cortex-wide neuronal activity in mice with high signal-to-noise ratio (SNR) and spatiotemporal resolution (Ren and Komiyama 2021; Nietz et al. 2022; Ma, Shaik, Kozberg, et al. 2016; Kozberg et al. 2016; Matsui, Murakami, and Ohki 2016; Ma, Shaik, Kim, et al. 2016). Given these capabilities, WFCI has been extensively employed to study the functional organization and the development of mouse brain during various states of consciousness (Brier et al. 2019; Wright et al. 2017; Zhang, Landsness, Culver, et al.

2022; Brier et al. 2021; Ma, Shaik, Kozberg, et al. 2016), behaviors (Allen et al. 2017; West et al. 2022), and disease models(Balbi et al. 2019; Cramer et al. 2019). More recently, WFCI has also been applied to characterize calcium dynamics of neural activity during sleep and proven as a useful tool to uncover the neural correlates of sleep (Turner et al., n.d.; Niethard et al. 2018; Niethard, Brodt, and Born 2021). In order to study the mouse brain physiology during sleep, it is necessary to score the spatiotemporal WFCI recordings by assigning individual epoch into various sleep-wake states such as wakefulness, non-rapid eye movement (NREM) and REM sleep. This procedure is traditionally conducted by manual inspection of adjunct electroencephalogram (EEG) and electromyogram (EMG) signals. However, this process is time-intensive, often suffers from low inter- and intra-rater reliability (Bliwise et al. 1984; Collop 2002; Danker-Hopfe et al. 2009; Drinnan et al. 1998; Lord et al. 1989; Loredo et al. 1999; Norman et al. 2000; Rosenberg Richard S. and Van Hout Steven, n.d.; Silber et al. 2007; Whitney et al. 1998) and requires invasive electrophysiology. To overcome these limitations, an automated sleep state classification method that operates on spatiotemporal WFCI data alone is desired.

We recently reported a hybrid, two-step method where multiplex visibility graphs and a convolutional neural network (CNN) were combined for sleep state classification from WFCI data (Zhang, Landsness, Chen, et al. 2022). However, this approach requires an additional mapping of raw WFCI data to graph representations and thus fails to fully explore the spatiotemporal nature of the WFCI data. In addition, the interpretability of spatial and temporal features exploited by the network that are relevant to sleep state classification is limited. Thus, a sleep state classification model that directly utilizes the spatiotemporal calcium dynamics recorded by WFCI holds potential to further improve the performance and interpretability.

Sleep is a cyclical process and thus, temporal dependencies could be important when inferring sleep state. One widely adopted approach in deep learning that is capable of learning order dependence of sequential data is the recurrent neural network, such as the long short-term memory (LSTM) network. Several recent applications have employed the LSTM to infer sleep states or categorize heart disease on sequential physiological signals such as EEG (Zhuang et al. 2022; Michielli, Acharya, and Molinari 2019; Mousavi, Afghah, and Acharya 2019) and electrocardiogram (ECG) (Cheng, Zou, and Zhao 2021; Saadatnejad, Oveisi, and Hashemi 2020). When assigning a sleep state to an individual epoch, it is typical for sleep experts to take into consideration adjacent time points and further locate the occurrence of discriminative features such as K-complexes, spindles and slow waves, which can guide them to make decisions (Buzsaki 2006; De Gennaro and Ferrara 2003; Halász 2016). Therefore, a bidirectional LSTM (BiLSTM) that can leverage temporal information from both past and future events could potentially benefit the sleep state classification performance. In addition, incorporating an attention mechanism into the model enables selective emphasis on different parts of sequential input with varying degrees of importance. This could potentially better leverage temporal signatures in the data that aligns more effectively with the identification of sleep states (Bahdanau, Cho, and Bengio 2016; Mousavi, Afghah, and Acharya 2019; Kwon et al. 2021).

In this work, we proposed a hybrid, two-stage deep neural network architecture consisting of a convolutional neural network (CNN) and a BiLSTM with attention mechanism to jointly learn spatial and temporal information from WFCI data for automated sleep state classification. The CNN was employed to extract spatial information from WFCI image frames and a BiLSTM module was applied to capture temporal dependencies between features extracted from each time points. To identify brain regions that contributes most to the decision of different sleep state, gradient-weighted class activation maps (Grad-CAM) was computed. In addition, the temporal characteristic of WFCI data for sleep state classification was investigated. We find that the proposed CNN-BiLSTM effectively identifies sleep states by use of the spatial and temporal neural dynamics provide by WFCI.

## 2. Methods

2.1 Mice and surgical preparations

A total of 16 transgenic mice (12-16 weeks of ages, all males) expressing GCaMP6f (JAX strain: C57BL/6J-Tg (Thy1-GCaMP6f) GP5.5Dkim; stock: 024276) in excitatory neurons were acquired from Jackson Laboratories and used in the experiments in this study. All studies were approved by the Washington University School of Medicine Institutional Animal Care and Use Committee and followed the guidelines of the National Institutes of Health *Guide for the Care and Use of Laboratory Animals*. Mice were housed in 12-hour light/dark cycles with lights on at 6:00 AM and given ad lib access to food and water. Prior to data collection, the head of each mouse was shaven, and a midline incision was made to expose the skill. A Pixelglass head cap was fixed with a translucent adhesive cement (C&B-Metabond, Parkell Inc., Edgewood, New York) to allow chronic, repeated imaging (Silasi et al. 2016). To implant EEG and EMG electrodes, copper EEG pins (Newark Electronics, catalog #89H8939) were placed at the surface (0.7 mm cranial burr holes) of the brain overlying the lateral somatosensory cortex (-0.7 mm posterior to bregma, and +4.5 mm lateral to bregma) and fixed with Fusio dental cement. A referenced EEG screw was placed on the surface of the cerebellum. To record muscle activity, two 23-gauge stainless steel needles were attached to the posterior aspect of the Plexiglass headcap and inserted bilaterally into the neck muscles.

2.2 Wide-field calcium imaging of sleep

Mice expressing GCaMP6f were placed in a black felt hammock with their head secured for one to three sessions ranging from 30 to 180 minutes until the EEG/EMG signal showed the presence of sleep. Once sleep was established, the mouse then underwent a 3-hour undisturbed WFCI session. All recordings occurred between 9:00 AM and 1:00 PM during the mice's normal sleeping hours in order to maximize the chance of recording sleep.

A cooled, frame-transfer EMCCD camera (iXon 897, Andor Technologies, Belfast, Northern Ireland, United Kingdom) overhead and four collimated LEDs were used for image acquisition, as previously described (Wright et al. 2017; Brier et al. 2019; 2021; Zhang, Landsness, Chen, et al. 2022). Sequential illumination was provided by the four LEDs: 454 nm (blue, GCaMP6 excitation), 523 nm (green), 595 nm (yellow), and 640 nm (red) for hyperspectral oximetric imaging. The LEDs were sequentially triggered at 16.8 Hz per channel. The field of view covered the dorsal surface of the brain resulting in an area of ~1 cm$^2$ with pixel resolution of approximately 78 $\mu$m x 78 $\mu$m.

2.3 Expert behavioral state scoring

Using the combination of the filtered EEG/EMG signal and spectrogram, sleep states of wakefulness, NREM and REM were manually assigned based on 10-second epochs by the author E.L., a certified sleep specialist with over 15 years of experience scoring sleep. In total, a number of 19,155 10-second epochs were collected for this dataset, including 12,674, 5,701 and 780 data epochs for wakefulness, NREM and REM, respectively. The dataset was randomly shuffled and split into 80%, 10% and 10% for training, validation and testing of the sleep state classification model.

2.4 Data processing

Prior to analysis, the WFCI raw data underwent the following image processing steps. First, to control for session-to-session changes in ambient light levels, the mean light levels of non-illuminated frames for the session was subtracted from each image frame. To account for the effect of photobleaching, all pixel time traces were detrended. The global signal was regressed from every pixel's time trace to remove global sources of variance, such as motion and cardiac pulsations. To correct for any fluorescent absorption-emission by hemoglobin and deoxyhemoglobin, a ratiometric correction was applied using the reflectance channels at the GCaMP6 emission wavelengths (523 nm LED) as a reference. Images were then spatially smoothed with a 5x5 pixel box Gaussian filter. To register images between mice to a common atlas space, images for each mouse were affine transformed to the Paxinos mouse atlas using the positions of bregma and lambda as reference. A white-light image of the skull for each mouse and region of interest was manually demarcated as either "brain" or "non-brain" mask. All subsequent analysis was performed on pixels labeled as "brain".

2.5 Attention-based CNN-BiLSTM network architecture and implementation details
2.5.1   2D Convolutional neural network to extract spatial features

The CNN extracted spatial information from each image frame using five convolutional blocks. Each CNN block includeed a 2D convolutional layer (Conv2D) comprised of 64 kernels of size of 3, followed by a max pooling (MaxPool) operation with a pool size of 2 and a stride of 2 and activated by the use of leaky rectified linear unit (LeakyReLU). The last convolutional layer was followed by a global average pooling (GAP) layer to minimize the risk of overfitting by reducing the number of parameters in the model. The CNN blocks were wrapped by time-distributed layers so that they can be applied to every image frame of the input. Figure *1* shows the illustration of the architecture employed in our study.

2.5.2   Bidirectional long short-term memory with attention mechanism to learn temporal dependencies

Long short-term memory is one kind of recurrent neural network that consists of memory blocks for capturing long-term dependencies of sequential data. Three types of gates are used in a LSTM unit: input gate, output gate and forgot gate, which control the flow of information (Hochreiter and Schmidhuber 1997). Different from uni-directional LSTM which is only able to leverage the information from previous events, bi-directional LSTM contains an additional LSTM layer that allows for the input sequence flows backward and the outputs from both LSTM layers are combined. The use of a bidirectional LSTM recurrent layer instead of a unidirectional one is motivated by the fact that both adjacent timepoints were inspected when human experts score sleep based on recorded neural activities. Therefore, temporal information from past and future time points could be beneficial to improve the classification performance. As illustrated in Figure *1*, given an WFCI data epoch as input to the spatial branch consisting of CNN blocks, feature vectors of each image frames are extracted. Subsequently, they are sent into the BiLSTM layer to learn the temporal dependency from both previous and following events. In this study, a number of 64 LSTM units were used for per LSTM layer.

Since sleep state scoring by human experts is typically based on the occurrence of temporal events such as K-complexes, spindles, theta rhythms and slow waves (Zhang, Landsness, Chen, et al. 2022; Buzsaki 2006; De Gennaro and Ferrara 2003; Halász 2016), the sequential feature vectors from the LSTM layers could contribute differently for the classification of different sleep stages. In order to improve the classification performance by allowing the LSTM network to focus on timesteps with more discriminative sleep-state related features, an attention module with Bahdanau's additive style (Bahdanau, Cho, and Bengio 2016; Tezuka et al. 2021) was added to learn the importance of feature vectors of each time points. Given the hidden state vector $\boldsymbol{h_i}$ at every time point $i$, the importance score $\boldsymbol{s_i}$ can be calculated by the score function as:

$$\boldsymbol{s_i} = \tanh(\boldsymbol{W_s h_i} + \boldsymbol{b_s}),$$

where $\boldsymbol{W_s}$ and $\boldsymbol{b_s}$ are trainable weights and bias, respectively.

The attention weight $\alpha_i$ was computed using the softmax function:

$$\alpha_i = softmax(s_i) = \frac{\exp(s_i)}{\sum_i s_i}$$

The final output $\boldsymbol{v}$

$$\boldsymbol{v} = \sum_i \alpha_i \boldsymbol{h_i}$$

Finally, a fully connected layer with softmax activation classifies the sleep state.

2.5.3   Implementation details

Each 10-s WFCI data epoch consisted of 168 image frames each in size of 128 X 128 and was given as input to the CNN-BiLSTM network. The network was trained by use of an Adam optimizer (Kingma and Ba 2017) with a learning rate of 0.0001 to minimize the focal loss (Lin et al. 2020), which is used to address the class imbalance of the sleep states labels. In our study, the focusing parameter γ = 2 was considered. The network was implemented in Python 3 with TensorFlow 2.2.0 using NVIDIA Quadro RTX 8000.

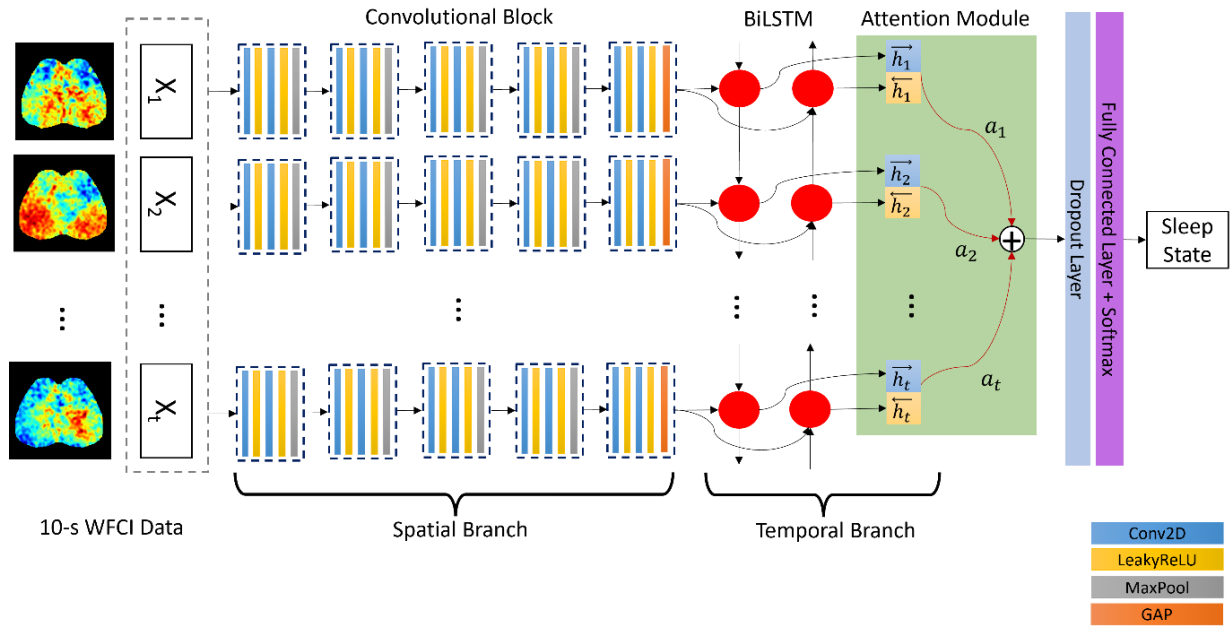

Figure 1 Illustration of attention-based CNN-BiLSTM model architecture used in the study. The CNN-BiLSTM is a hybrid model consisting of a spatial branch to extract spatial features of WFCI image frames and a temporal branch to capture the temporal dependency among time steps in the WFCI recordings. Five convolutional blocks (within black dashed line) were employed in the CNN, in which two convolutional layers (blue) activated by Leaky ReLU (yellow) and max pooling layer (gray) were included. A global average pooling layer (orange) is placed after the final convolutional layer. The following bidirectional LSTM architecture receives the extracted spatial features from CNN as inputs. The BiLSTM layer (red) is composed of 64 units in each direction. Temporal attention mechanism is applied to the BiLSTM outputs before they are fed to a fully connected later with softmax activation to perform the final classification.

2.6 Performance evaluation

The models were evaluated on test data consisting of unseen epochs from the same group of training subjects as well as on an independent subject. Metrics including precision, accuracy and F1-score were utilized to evaluate the model performance. The Cohen's kappa statistic, κ (Cohen 1960), was computed to assess the inter-rater reliability between manual EEG/EMG-based scoring and the proposed automated CNN-BiLSTM classification results. A kappa magnitude between 0.61 and 0.80 indicates a substantial agreement between the two raters (McHugh 2012). A confusion matrix for each classification task was formed to provide a direct interpretation of the classification results.

2.7 Data and source code availability

Part of the the WFCI sleep data is available on PhysioNet (Landsness, Eric and Zhang, Xiaohui n.d.; Goldberger et al. 2000). All model training and testing code are available at https://github.com/comp-imaging-sci/attention-based-bilstm-sleep-scoring.

## 3. Results
3.1 Attention-based CNN-BiLSTM classifies sleep states as wakefulness, NREM and REM

To automatically classify sleep states as wakefulness, NREM, and REM, CNN-BiLSTM was trained and tested on spatiotemporal WFCI data (Figure 1). The sleep state classification results of the CNN-BiLSTM on WFCI alone were compared to human-scored EEG/EMG that were simultaneously collected with the WFCI data to assess the performance. For the individual sleep states in the test set consisting of 10% of the data, the precision (recall) was 0.89 (0.88) for wakefulness, 0.71 (0.74) for NREM and 0.83 (0.68) for REM (Table 1, Figure 2). The CNN-BiLSTM achieved an F1-score of 0.84 and Cohen's κ value of 0.64, where a κ value > 0.60 is indicative of substantial agreement (Table 1). In addition, the WFCI data was mapped to multiplex visibility graphs (MVG) and a compact 2D CNN was trained on the MVG representations to compare the classification performance of the MVG-CNN (Zhang, Landsness, Chen, et al. 2022) to the proposed CNN-BiLSTM. As can be seen in Table 1 and Figure 2, the performance achieved by CNN-BiLSTM is comparable to that of MVG-CNN.

Table 1 Metrics to evaluate the sleep stage classification performance on test data (n=1810 epochs). Both WFCI-based CNN-BiLSTM model and MVG-CNN model achieved substantial agreement of $\kappa = 0.64$ compared to manual EEG/EMG-based scoring. Prec., precision; Rec., recall; Acc., accuracy; $\kappa$, Cohen's Kappa statistic.

| Scoring method | Wakefulness | | NREM | | REM | | Acc. | F1-score | κ |
|---|---|---|---|---|---|---|---|---|---|
| | Prec. | Rec. | Prec. | Rec. | Prec. | Rec. | | | |
| **CNN-BiLSTM** | 0.89 | 0.88 | 0.71 | 0.74 | 0.83 | 0.68 | 0.83 | 0.84 | 0.64 |
| **MVG-CNN** | 0.87 | 0.91 | 0.74 | 0.72 | 0.93 | 0.60 | 0.84 | 0.84 | 0.64 |

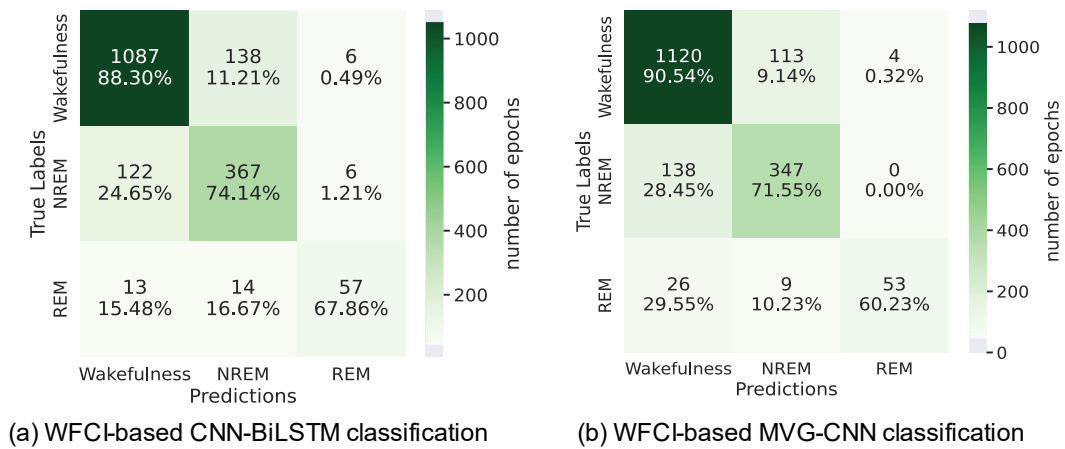

(a) WFCI-based CNN-BiLSTM classification    (b) WFCI-based MVG-CNN classification

Figure 2 Confusion matrix for the CNN-BiLSTM model and the MVG-CNN model on sleep classification of wakefulness, NREM, and REM in test set (n=1810 epochs), respectively. Manual EEG/EMG-based scoring is on the x-axis and MVG-CNN predictions are on y-axis. The diagonal cells correspond to the numbers of correctly classified epochs and precision rate (%) across wakefulness, NREM and REM states. Non-diagonal cells indicate misclassified epochs for each state.

To further demonstrate the ability of the CNN-BiLSTM to classify sleep states from spatiotemporal WFCI data, sleep states of an unseen 3-hour WFCI recording were classified. The CNN-BiLSTM achieved a κ of 0.67, indicating a substantial agreement between EEG/EMG-based scoring and the CNN-BiLSTM classification, and comparative to our previous MVG-CNN approach. In order to further compare EEG/EMG-based scoring and CNN-BiLSTM classification, we analyzed measures of sleep fragmentation, sleep-state organization and spectral power. While the CNN-BiLSTM method caused shorter sleep state durations and an increased number of state transitions compared to the EEG/EMG-based method (Table 2), it has less state transitions compared with our previous MVG-CNN. This suggests the potential benefit of LSTM to better capture temporal dependency than use of MVGs.

Table 2 Comparison of sleep fragmentation among WFCI-based sleep state classification by use of the proposed CNN-BiLSTM model, the MVG-CNN method and the EEG/EMG-based human scoring.

| Scoring method | Average sleep state length (sec) | | | Number of state transitions |
| --- | --- | --- | --- | --- |
| | Wakefulness | NREM | REM | |
| CNN-BiLSTM | 35 | 57 | 37 | 238 |
| MVG-CNN | 29 | 61 | 35 | 245 |
| Human annotator | 47 | 66 | 81 | 182 |

As depicted by the hypnogram (Figure 3a, c, e), there was substantial agreement in the temporal pattern (sleep cycles) of transitions between wakefulness, NREM and REM. In addition, both EEG/EMG scored by a human and WFCI classified by the CNN-BiLSTM showed an increase in delta (0.4–4.0 Hz) spectral power of the calcium signal exclusive to NREM and an increase in theta (6.0–8.0 Hz) exclusive to REM (Figure 3b, d, f), confirming the effective classification of sleep states by both methods. Further, this agreement between EEG/EMG-based human scoring and WFCI-based CNN-BiLSTM classification is comparable to the inter-rater reliability of two human experts with a κ of 0.65 (Supplemental Figure 1). These results confirms that sleep states classified by the CNN-BiLSTM using WFCI data alone are highly similar to EEG/EMG-based human scoring.

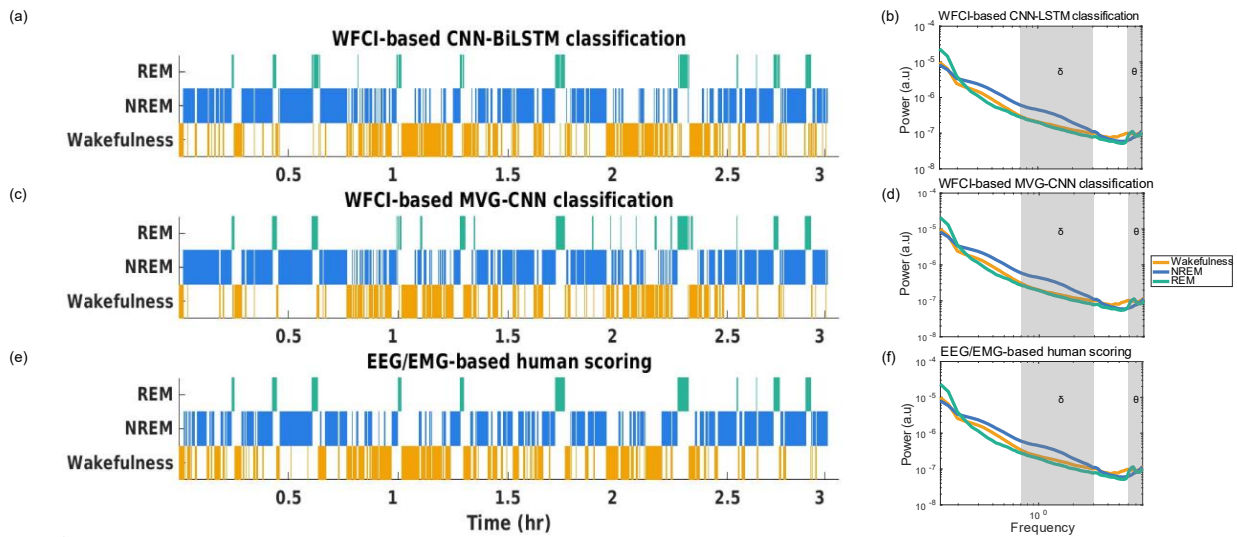

Figure 3 Comparison of sleep state classification among CNN-BiLSTM, MVG-CNN and human annotator on a 3-hour recording of a mouse. Hypnograms corresponding to (a) CNN-BiLSTM

classification based on WFCI recording, (c) MVG-CNN classification based on WFCI recording and (e) human EEG/EMG-based scoring. Average power spectra of the calcium signal plotted for wakefulness, NREM and REM based on the (b) predictions from CNN-BiLSTM, (d) predictions from MVG-CNN and (f) true scoring produced by a human annotator. Shaded gray areas represent the delta (δ, 0.4-4.0 Hz) and theta (θ, 6.0-8.0 Hz) frequency ranges.

3.2 Attention-based CNN-BiLSTM locates spatial characteristics for sleep state classification

There is emerging evidence of spatially unique patterns of neuronal activity under different states of sleep (Dong et al. 2022; Wang et al. 2022; Bréchet et al. 2020). Here, Grad-CAM (Selvaraju et al., 2020) was computed to employ class-specific gradient information flowing into the final convolutional layer of a CNN to produce a coarse localization map of regions of emphasis. When applied to the spatiotemporal WFCI data, Grad-CAM identified brain regions that contribute most to a classification decision and revealed different patterns of emphasis for various sleep states (Figure 4). For example, in NREM state, the frontal brain regions such as motor and somatosensory cortex received more attention from the network while the posterior regions including visual and retrosplenial cortex were emphasized more. These observations also hold when only hemispheric data was given as training input (Supplemental Figure 2). Taken together, these results show the CNN-BiLSTM classifies NREM and REM sleep based on anterior and posterior areas, respectively.

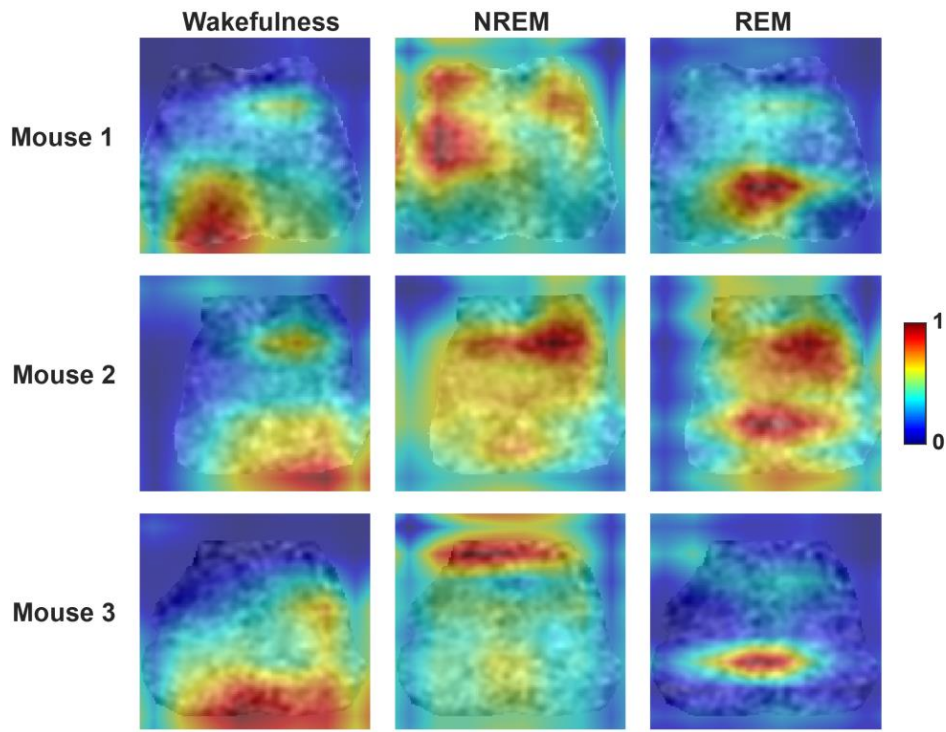

Figure 4 Representative Grad-CAM examples of wakefulness, NREM and REM from three mice. A higher intensity with the color gradients (i.e., red, value 1) reveals that the 2D CNN focuses more on such regions of interest when making corresponding decisions.

3.3 Attention-based CNN-BiLSTM reveals temporal characteristics for sleep state classification
3.3.1 MVG-CNN uses short- and long-range temporal information to classify sleep

In order to understanding how the CNN-BiLSTM leverages the temporal information of WFCI to classify sleep-wake states, the network was trained on 20-s data along with the ground truth label of the central 10 s. The goal of this study was to investigate the importance of temporal information in various range within the WFCI data in different states. The weights from the attention layer were extracted for 20-s epoch and visualized for wakefulness, NREM and REM (Figure 5). It was found that the network focuses on long-range temporal dependency when making decision to classify the input as REM states. In contrast, NREM focuses on short-range temporal information and larger weights are potentially given to the slow oscillations in the recorded neural activity.

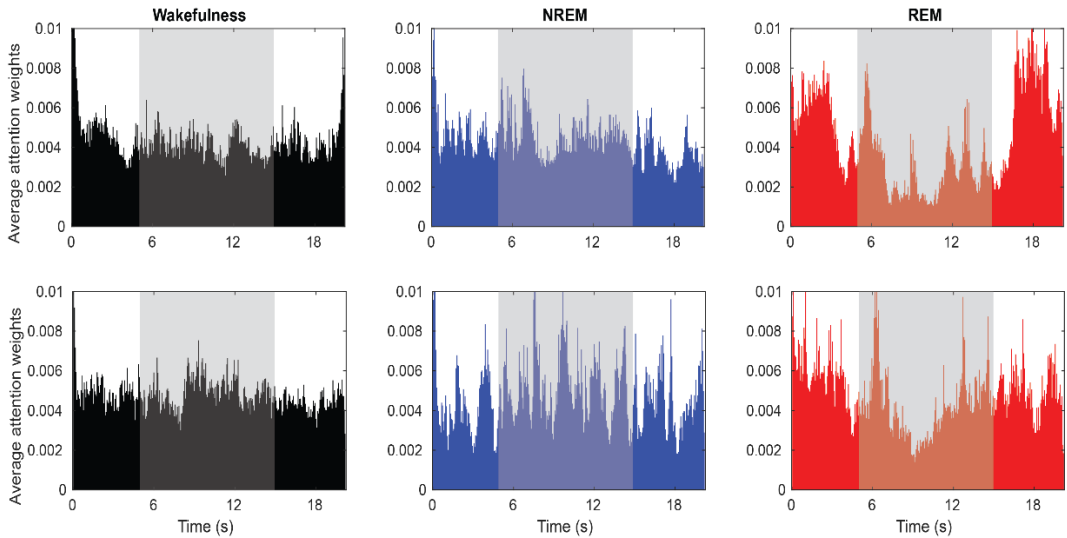

Figure 5. Attention weights for 20-s data epoch of wakefulness (black), NREM (blue) and REM (red).

3.3.2   Varying epoch duration impacts sleep state classification performance

Human experts conventionally score sleep EEG/EMG signals from mice with an arbitrary 10-second epoch duration. However, human-defined sleep epochs contain a mixture of sleep states with the predominant state being classified. This mixture of states raises the question of whether shorter epoch durations would lead to better sleep state classification performance of WFCI data (Yan et al., 2011). Here, we varied the epoch duration from 1 to 20 s to investigate the impact of temporal information incorporated from spatiotemporal WFCI data on sleep state classification performance (Figure 6). As the epoch duration was increased from 1 s to 10 s, the classification accuracy and Cohen's κ improved. At an epoch duration of 10 s and higher, accuracy plateaued at ~0.84 with Cohen's of ~0.64. These results suggest that shortening epochs below 10 s or increasing beyond 10 s may not benefit sleep state classification performance by the CNN-BiLSTM for the WFCI dataset being classified.

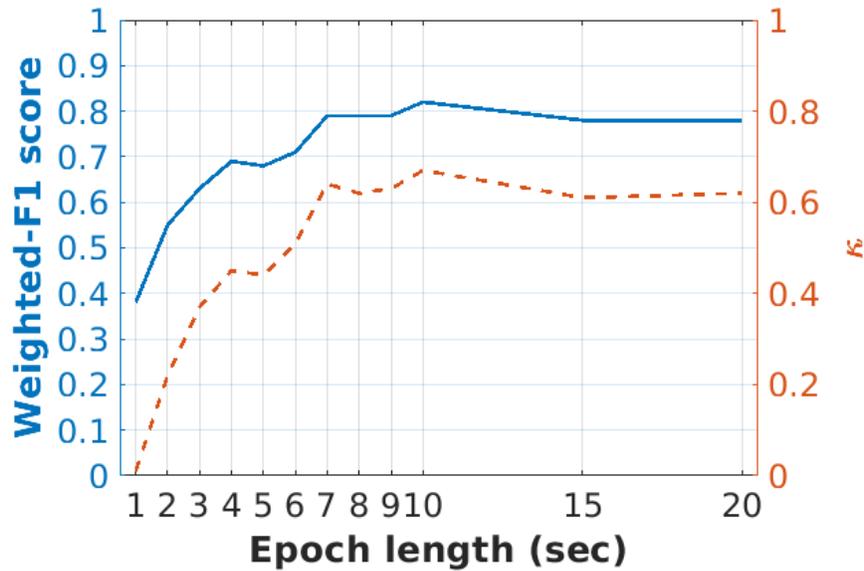

Figure 6 The sleep state classification performance with respect to the epoch durations used in the CNN-BiLSTM. As epoch duration was varied, weighted F1-score (left y-axis, solid blue line) and the Cohen's Kappa statistic (right y-axis, dotted red line) were compared.

## 4. Discussion

In this study, we proposed a hybrid, two-stage deep neural network architecture consisting of a convolutional neural network and a bidirectional long short-term memory with attention mechanism to accurately classify sleep states from spatiotemporal WFCI data alone. Leveraging spatial information, we identified frontal and posterior regions being important for classifying WFCI data as NREM or REM sleep respectively. Through the use of attentional weights, showed a temporal dependency in network decision making with REM sleep relying on relatively long-scale dependencies. The improved interpretability of the proposed network allows us to gain deeper insights into the spatiotemporal calcium dynamics that underlies sleep.

Automated accurate sleep state classification methods are desired for sleep research with WFCI. While recent research that addresses automated sleep scoring has attracted increasing attention, most of the proposed algorithms primarily work with biosignals such as photoplethysmograms (Korkalainen et al. 2020; Wu et al. 2020), heart rate and movement, as well as electrophysiological signals from EEG and EMG. Recent advances in wide-field calcium imaging enables researchers to record high-dimensional neural activities across various states such as spontaneous wakefulness, NREM, and REM sleep in mice (Niethard et al. 2016). In this context, we have successfully implemented a CNN-BiLSTM model to distinguish sleep states directly from spatiotemporal data acquired through WFCI. The proposed CNN-BiLSTM demonstrated performance on par with the established gold standard of inter-rater reliability among expert human EEG/EMG scorers (Rosenberg Richard S. and Van Hout Steven, n.d.). While it does not surpass the performance of our previous MVG-CNN method (Zhang, Landsness, Chen, et al. 2022), its ability to process raw spatiotemporal WFCI data enhances the interpretability of the temporal and spatial features of calcium dynamics during sleep. Thus, our hybrid CNN-BiLSTM method is an effective and accurate tool for automatic sleep scoring that also offers greater interpretability.

Sleep is a temporal process with strong dependency between consecutive time points (Zhao, Li, and Guo 2022). When human experts assess sleep patterns, they take into account both the preceding and subsequent events to make decision regarding specific sleep states. In this study, we sought to emulate this decision-making process by incorporating a bidirectional LSTM network and thus fully leverage temporal dependencies in both forward and backwards directions of sleep dynamics captured by WFCI. In addition, when analyzing temporal neural activity, it is of interest to understand which time point holds greater significance than the others, which allows us to further investigate the potential markers of a sleep event. Here, the attention weights extracted from our network reveals unique patterns for classifying different sleep states. For instance, the CNN-BiLSTM characterized REM sleep by focusing on long-scale dependency, aligning with observation from our prior MVG-CNN method (Zhang, Landsness, Chen, et al. 2022). In contrast, during NREM, the network exhibited a heightened emphasis on low frequency components, potentially corresponding to the slow waves observed in NREM (Niethard, Brodt, and Born 2021; Brier et al. 2019). Future studies investigating the signatures of EEG/EMG signals that align with high attention scores could enhance our understanding about the underlying mechanisms governing sleep events.

Sleep is not a unitary, homogeneous state but exhibits spatial diversity across the cortex. A major question within the field of sleep field is whether sleep can occur in confined regions of the cortex. Our findings indicate that frontal areas play a pivotal role in the identification of NREM sleep, aligning with the observation of strong inhibition of frontal pyramidal neurons (Li et al. 2023). Conversely, posterior regions seem to be more engaged during REM and wakefulness. This distinction in sleep state classification highlights the spatial heterogeneity of sleep, confirming that sleep is not a unitary, homogeneous state but is spatially diverse across the cortex(Zhang, Landsness, Chen, et al. 2022). Future WFCI studies while leveraging techniques such as optogenetics will allow for better understand the sleep mechanism in these localized brain areas.

To our knowledge, this study is the first to develop a deep learning-based automated sleep state classification method that directly works on spatiotemporal WFCI data. The proposed method enhances the interpretability of the network's decision-making process in assigning specific sleep state. However, there are possible iterative improvements for the classification of WFCI sleep-wake states. For example, the use of a transformer network that can model long-range dependencies and offer faster computational speed could further our understanding of the spatiotemporal characteristics in various sleep states (Wan et al. 2023; Du et al. 2022). Additionally, the application of diffusion models with high-quality image generation to augment imbalanced sleep data could potentially improve classification performance (Zhang et al. 2023).

## 5. Conclusion

In this study, we described an automated sleep state classification method that operates directly on spatiotemporal WFCI data. Utilizing a convolutional neural network, our approach learns the spatial features of image frames while a bidirectional LSTM with an attention mechanism captures the temporal dependencies across different time points. The proposed CNN-BiLSTM model achieved substantial agreement with manual EEG/EMG-based scoring and provides enhanced interpretability of the networks' decision-making in categorizing various sleep states. We uncovered the importance of spatial information across brain regions and temporal characteristics of calcium dynamic, respectively. Our findings advocates for the use of CNN-BiLSTM in elucidating the neural correlates of sleep via spatiotemporal WFCI and demonstrated its potential for broader research applications.

**Acknowledgements**

This work was supported in part by the National Institute of Neurological Disorders and Stroke (R01NS099429 to J.P.C., R37NS110699 and R01NS094692 to J.M.L., K08NS109292-01A1 to E.C.L.), National Institute on Aging (F30AG061932 to L.M.B.), American Academy of Sleep Medicine Foundation (201-BS-19 to E.C.L.) and American Heart Association (20CDA35310607 to E.C.L.).

**Disclosure Statement**

The authors declare no potential conflicts of interest.

**Supplementals**

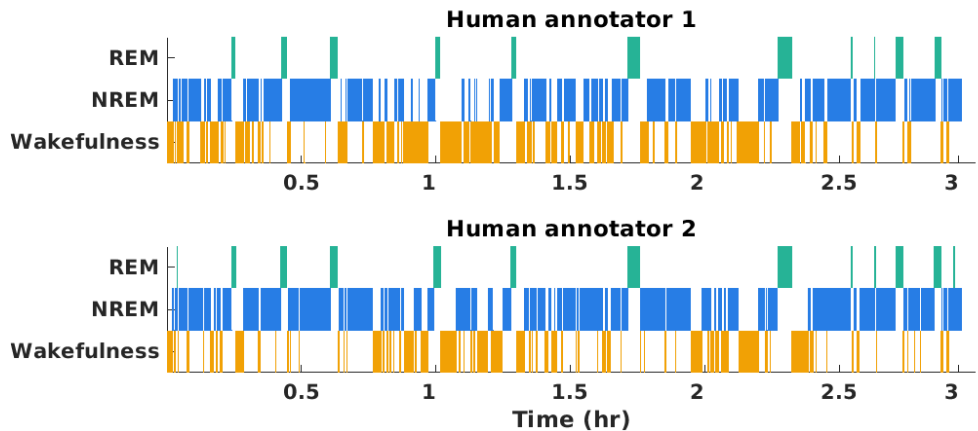

Supplemental Figure 1 Comparison of sleep state classification between two human annotators on a 3-hour recording of a mouse. Cohen's κ is 0.65.

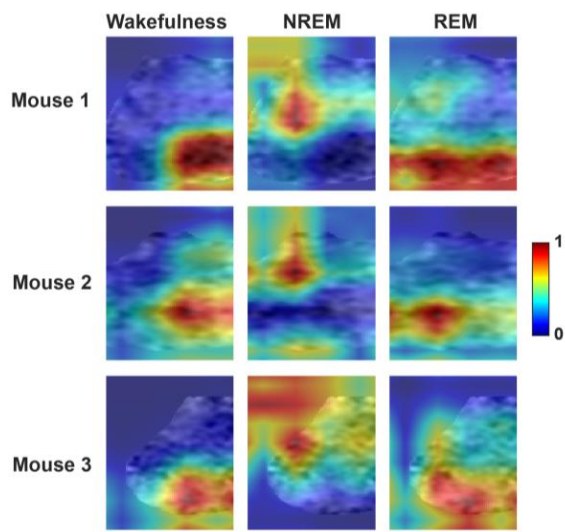

Supplemental Figure 2 Representative Grad-CAM examples of wakefulness, NREM and REM of hemispheric data from three mice. A higher intensity with the color gradients (i.e., red, value 1) reveals that the 2D CNN focuses more on such regions of interest when making corresponding decisions.